# Surface passivation and oxide encapsulation to improve optical properties of a single GaAs quantum dot close to the surface

Santanu Manna*[a], Huiying Huang[a], Saimon Filipe Covre da Silva[a], Christian Schimpf[a], Michele B. Rota[b], Barbara Lehner[a], Marcus Reindl[a], Rinaldo Trotta[b] and Armando Rastelli[a]

[a]*Institute of Semiconductor and Solid State Physics, Johannes Kepler University, Linz 4040, Austria*
[b]*Department of Physics, Sapienza University of Rome, 00185 Rome, Italy*

*Corresponding author: santanu.manna@jku.at

**Abstract**

Epitaxial GaAs quantum dots grown by droplet etching have recently shown excellent properties as sources of single photons as well as entangled photon pairs. Integration in some nanophotonic structures requires surface-to-dot distances of less than 100 nm. This demands a surface passivation scheme, which could be useful to lower the density of surface states. To address this issue, sulphur passivation with dielectric overlayer as an encapsulation is used for surface to QD distances of ≲ 40 nm, which results in the partial recovery of emission linewidths to bulk values as well as in the increase of the photoluminescence intensity.

## 1. Introduction

Single photon emitters are regarded as key elements in quantum optical networks for transferring qubits between remote network nodes. Solid state semiconductor quantum dots (mainly GaAs- and InP-based) are arguably the most promising sources for the integration into a practical and scalable networks [1–7]. However, in comparison with single atoms and ions in traps, quantum dots (QDs) are subject to more pronounced decoherence and spectral diffusion [8–10]. Out of those, the spectral diffusion, which determines random changes in the QD optical emission frequencies, becomes a big issue whenever a QD sits close to the surface ($\lesssim$ 100 nm for GaAs-based QDs) [9,11,12]. This is due to the presence of a large density of surface states (e.g. $(12.5\pm10.0)\times10^{13}$ eV$^{-1}$cm$^{-2}$ at 300 K for GaAs [13]) at GaAs surface. A systematic decrease of emission linewidth with increasing QD-surface distance has been consistently observed by several groups [11,12,14]. Also, other quantum optical properties such as single photon emission probability and Hong-Ou-Mandel visibility are expected to benefit from a large separation between QD and surface. For many applications it is necessary to embed the QDs in nanophotonic structures like planar antenna [15,16], nanopillar [17], nanocone [18], circular Bragg grating structure [19–21] etc., where QD to surface distance of <100 nm is really essential to out-couple the dot emission efficiently from a high refractive index material like GaAs. Therefore, a reduction of the surface state density is important to reduce charge fluctuations and hence spectral diffusion. To significantly reduce the surface state density proper surface passivation technique could be implemented. A specific example of photonic structure which would benefit from passivation is a planar Yagi-Uda antenna [16] where, for GaAs QDs operating at wavelengths around 780 nm and/or InGaAs QDs around 900 nm, it is necessary to have the dot to surface distance ≤ 40 nm for efficient outcoupling. Many works have been performed over several decades in order to get a proper passivation of III-V compound semiconductor surfaces mostly for electronics devices [22–28]. Such methods have been recently employed also for improving the performance of optical devices [11,22,29–32].

The aim of this work is to assess the effect of surface passivation with a dielectric over-layer on the emission linewidth and intensity of GaAs QDs located a few tens of nm ($\lesssim$ 40nm) away from the exposed surface.

## 2. Material and methods

### 2.1. Growth and processing

In this work, GaAs QDs embedded within an $Al_xGa_{1-x}As$ matrix (in this case x=0.2 or 0.4) were epitaxially grown using the droplet etching method [33,34] on a GaAs (001) substrate. It is already known that this method allows creating close-to-symmetrical dots with very low excitonic fine structure splitting (FSS) [35], which is critical for polarization entangled photon generations with near-unity entanglement fidelity [36]. Here we concentrate on the linewidth of the excitonic emission of single QDs. We studied two kinds of structures (see Fig. 1) : (i) Str-1: In the 'As-grown' structure, the distance between the QD layer and the 1$^{st}$ GaAs cap layer is 36 nm and that to the very top surface is 126 nm; we then selectively etched the top GaAs cap layer and the underlying $Al_{0.75}Ga_{0.25}As$ sacrificial layer (the result is indicated with 'Only etched' in Fig. 1); finally we applied the surface passivation steps, which include sulphur passivation and $Al_2O_3$ deposition, to obtain a 'Passivated' sample; (ii) Str-2: in the 'As-grown' structure, the QD-to-surface distance is 72 nm; for 'Only Etched', the GaAs cap is etched and also some portion of the $Al_{0.2}Ga_{0.8}As$ matrix, so that the QD to surface distance is 40 nm without any GaAs cap; for 'Passivated', the sulphur passivation scheme is followed by $Al_2O_3$ deposition on the bare $Al_{0.2}Ga_{0.8}As$ surface.

Etching of GaAs or $Al_{0.2}Ga_{0.8}As$ was performed with a mixture of citric acid and $H_2O_2$ with a ratio of 25:1, with roughly the same etching rate of ~2.5 nm/sec [37]. The $Al_{0.75}Ga_{0.25}As$ sacrificial layer for Str-1 is etched selectively using a buffered oxide etch (BOE) HF with an etching rate of ~5 nm/sec [38]. After immediate etching of the sacrificial layer, HCl treatment is done to remove native oxide by dipping in $HCl:H_2O$ (1:1) solution for 1 min followed by deionized water rinsing and $N_2$ drying. Afterwards, $(NH_4)_2S$ treatment is accomplished by soaking the sample in $(NH_4)_2S$ for 10 min at room temperature followed by $N_2$ drying [39]. For some samples (with Str-1), we carried out an alternative treatment involving $NH_4OH$ instead of $(NH_4)_2S$ [39]. It was done by soaking the samples in 20% $NH_4OH$ solution for 3 min followed by rinsing in flowing deionized (DI) water and $N_2$ drying. After these surface pretreatments, the samples were immediately loaded inside an atomic layer deposition (ALD) chamber to deposit $Al_2O_3$ layer by using alternating pulses of trimethyl aluminum (TMA) and water ($H_2O$) precursors at a substrate temperature of 200 ºC. Two different $Al_2O_3$ layer thicknesses of 6 and 12 nm were used for both sample structures.

*2.2. Optical measurements*

The optical measurements were carried out using a micro-photoluminescence (μ-PL) setup. A continuous flow helium cryostat was used to cool down the samples to 5 K. A continuous wave (cw) laser at 514 nm was used to excite the GaAs QDs. An objective lens with a numerical aperture of 0.65 (spot size of ~1 μm in diameter) was used to focus the laser onto the samples and also to collect the PL of the QDs with a typical integration time of 1 sec. The PL signal was sent through a monochromator and detected with a liquid-nitrogen-cooled CCD. The resolution of our monochromator is approximately 40 µeV in the QD emission wavelength range (around 780 nm).

## 3. Results and discussion

*3.1. Role of passivation/oxide encapsulation*

We now discuss the function of the sulphur passivation and oxide encapsulation. Unlike native $SiO_2$ on Si, native oxides of GaAs possess built-up charges at the surface/interface [13]. $(NH_4)_2S$ is employed to remove the native oxides of GaAs, if any, followed by the etch-down of GaAs by a few Å and incorporation of S-atoms to saturate the GaAs surface dangling bonds [40]. Henceforth, a lower probability holds for GaAs surface to get oxidized. In spite of that, the laser illumination during photoluminescence measurements can make it possible to change Ga-S, As-S bonds to Ga-O and As-O and so also to increase again the surface state density [41]. For this reason we protect the S-passivated GaAs surface with an $Al_2O_3$ capping layer obtained by ALD. This method is a slow monolayer-by-monolayer deposition technique based on surface chemical reactions, which can provide high-quality, thermodynamically stable oxides on III–V semiconductors like GaAs [39]. In addition to that, ALD growth for $Al_2O_3$ works as a self-cleaning reaction process for an extra removal of native oxides after $(NH_4)_2S$ treatment [40]. Here, the dielectric $Al_2O_3$ layer acts as a diffusion barrier to prevent re-oxidation of the semiconductor region.

*3.2. Role of band bending and dot optical properties*

In Fig. 2 we show the schematic band bending and different dot positions with respect to the surface for a p-type semiconductor for different surface conditions. Since the background doping in our samples is p-type, the band bending is similar like Fig. 2. However, the band bending could be in the opposite direction from the one shown in that figure in case of an n-type semiconductor [13,26]. In an ideal case of no surface states band bending does not occur in the bulk or near the surface (Fig. 2(a)). But, with the formation of surface states, a negative space charge region layer (depletion mode) will be induced in the GaAs cap–$Al_xGa_{1-x}As$ layer near the surface by the occupation of the surface states by holes, resulting in the bending of band edges downwards near the surface (Fig. 2(b)-(c)). A small density of surface states results in a small band bending (Fig. 2(c)), whereas in the case of a high density of surface states (Fig. 2(b)) (more than one surface state per hundred surface atoms [42]), the surface states are filled to a level close to the surface Fermi level and the band bending is significant and Fermi level pinning is likely. In our case, 'Only etched' samples without any kind of passivation exhibit similar nature like Fig. 2(b), whereas after passivation it is like Fig. 2(c). It is noteworthy that the electric field, arisen out of the band bending driven potential gradient, drops linearly when moving away from the surface to the end of the space charge region and this electric field is of fluctuating character because of trapping/detrapping of charge carriers on the surface. Therefore, a QD located within the depletion region and close to the surface will feel more fluctuating environment compared to the position near the end of the depletion region or beyond it. As a result, the exciton emission energy will be influenced depending on the position of the QDs. Since the dependence of the DC Stark energy shift ($\delta E$) on electric field ($F$) is quadratic ($\delta E \propto F^2$), the magnitude of the change in energy shift [$\delta(\delta E)$] and so also the linewidth, owing to a fluctuating field ($\delta F$), will depend on the average field as well as the fluctuating field [$\delta(\delta E) \propto F \delta F$]. Another possible scenario is that the electric field present in the depletion region polarizes the exciton inducing an electric dipole, which then interacts with fluctuations in the electric field or nearby charges more strongly than an unpolarized exciton. Consequently, QDs far away from surface, with a smaller average field, should possess narrower diffusion-induced linewidths than the one close to the surface. Hence the sulphur passivation scheme along with the encapsulation is helpful to reduce the density of surface states (by more than one order of magnitude [43]), leading to the reduction in the band bending, and so also the line broadening. It is noteworthy that the surface passivation can only decrease the surface states density but cannot change anything in the bulk region. A local electric field created by charge carrier trapping/detrapping near a QD might also contribute to the spectral diffusion or PL intensity dropping [9,44,45].

*3.3. Optical measurements*

In Fig. 3 we present low temperature (~5 K) PL spectra for arbitrarily selected single QDs in the different structures illustrated in Fig. 1. Neutral excitons along with different charge complexes are typical in the spectra. Among all the spectra, 'Only etched' case for Str-2 show the undulated background over all the powers ranging from the start of QD emission to its saturation. This might be attributed to the detrimental surface etching and the absence of GaAs cap, where the AlGaAs matrix tends to be converted into oxide over time. The propagation of oxide, and hence the appearance of background are mitigated to some extent via passivation with a thicker $Al_2O_3$ capping layer (in case of 'Passivated' samples for Str-2). For all the structures, the as-grown samples show resolution limited linewidth, whereas in case of 'Only etched' samples the linewidth significantly increases and partial recovery of linewidth is possible only after passivation. Fig. 4 illustrates representative spectra of the neutral exciton confined in QDs in the different structures accompanied by the corresponding fits. The lineshape fitting could be best performed by a Gaussian function instead of a Lorentzian one. A detailed measurement on line broadening using photon correlation Fourier spectroscopy (PCFS) has been put forth in same kind of droplet etching epitaxy grown GaAs dots embedded in a distributed Bragg reflector cavity (dot-to-surface >200 nm) by Schimpf *et al.*[46] from the same lab, which has demonstrated that from millisecond delay time scale onwards the lineshape shows inhomogeneous character. Hence it makes sense to consider the same also for our case. Fig. 5 shows bar plots for the linewidths extracted by Gaussian fitting of neutral exciton spectra of 30 different QDs from each sample. The error bar reflects the standard deviation of the measurements, reflecting statistical fluctuations. For both structures, the linewidth is resolution limited for as-grown samples, whereas upon etching and making the surface closer to the dots, the linewidth increases significantly to 63±7 µeV and 84±16 µeV for Str-1 and Str-2, respectively, but it reduces again after sulphur passivation. To test whether the improvement is really due to the Sulphur treatment, we replaced that with a $NH_4OH$ treatment after etching the cap layer of Str-1. This treatment makes GaAs surfaces -OH terminated and expected to decrease the surface states [39]. However, no improvement in PL linewidth decrement can be observed (inset of Fig. 5) by this treatment, although improvement in electrical properties have been shown before [39]. This might be due to the laser illumination which rapidly degrades the surface and increases density of surface states [41].

Along with the line broadening, a slight decrement of integrated PL intensity takes place for the 'Only etched' cases compared to the 'As-grown' or 'Passivated' ones, as shown in the Fig. 6. This comparative PL intensity study is performed by saturating the neutral exciton emission intensity. In this case, the decrement in intensity is not large (<1.5), as shown before, for a surface to dot distance of ~40 nm [12]. Furthermore, it was shown in [12] that a distance of <15 nm can cause a substantial reduction in PL intensity by more than 50%. In the case of Str-2, a 12 nm thick $Al_2O_3$ is found to be more functional than 6 nm $Al_2O_3$ capping to recover the intensity, and it might be due to the absence of the GaAs cap layer and detrimental surface etching. Since a QD inside the space charge region is strongly influenced by the electric field, the polarization of exciton can take place, which can lead to the decrease in the electron-hole overlap and so also the PL intensity. Another effect can stem from the fact that the excitation energy is larger than the bandgap of the barrier. Tunneling of captured carriers from quantum dot to the surface as well as carrier diffusion to the surface from the barrier region can take place. Both will lead to the decrease of PL intensity. Consequently, PL intensity will increase with increasing top tunnel barrier thickness as well as increasing the barrier height. One could also use excitation energy much lower than the barrier bandgap to avoid the carrier diffusion [14].

*3.4. Analysis of random Gaussian fluctuation*

In this section we estimate the strength of the random fluctuating environment, a characteristic figure responsible for the amount of spectral diffusion, which in turn is dependent on the density of surface states. Considering a system coupled to a fluctuating reservoir, we can extract the fluctuation strength or amplitude by employing Kubo-Anderson's stochastic theory [47–49]. In our case the origin of this random fluctuation process is nothing but the trapping/detrapping of charge carriers in presence of laser excitation at the surface. Assuming a Gaussian fluctuation of Markovian character, the lineshape function $I(\omega)$ takes the form [47]:

$$I(\omega) = \frac{2}{2\pi} \mathrm{Re} \int_0^\infty \exp\left\{-\frac{\Delta^2}{\gamma^2}\left(e^{-\gamma t} - 1 + \gamma t\right)\right\} \exp(-i\omega t)\, dt, \qquad (1)$$

where $\Delta$ is the amplitude or strength of the random modulation, $\gamma$ is the inverse of the correlation time or relaxation frequency and $C(t) = \exp\left\{-\frac{\Delta^2}{\gamma^2}\left(e^{-\gamma t} - 1 + \gamma t\right)\right\}$ is the correlation function. Depending on the strength of the random modulation of the system's frequency, the lineshape could be Gaussian or Lorentzian or a mixture of those [47]. To explicitly achieve that we have performed the Fourier transform of the correlation function $C(t)$ without any approximation using Mathematica package directly for different values of $\alpha\,(=\gamma/\Delta)$. It comes out that if the amplitude of the random modulation is stronger than the relaxation frequency, the lineshape becomes dominated by the Gaussian nature, whereas reducing the amplitude with respect to the relaxation pushes it towards the Lorentzian nature. This implies to the validity of slow modulation or inhomogeneous limit for our case and the lineshape happens to be the direct reflection of the random distribution of the emission frequencies. Considering strong random modulation we have calculated and plotted the values of $\Delta$ (linewidth $\sim 2\Delta\sqrt{2\ln 2}$) for different samples and shown it in Fig. 7. It is clear from the Fig. 7 that the fluctuation strength increases for the 'Only etched' cases, as the broadening takes place for the excitonic line, and can be suppressed with the help of passivation via decreasing surface states. In case of 'Only etched' samples with Str-2, $\Delta$ is more dispersed compared to other cases. It is to be noted that the lineshape looks almost Gaussian for $\alpha < 0.3$ (checked with Mathematica) and so the relaxation frequency follows $\gamma < 0.3\,\Delta$. Hence, we can extract the lower bound of the correlation time of charge trapping/detrapping as 400 ±72 ps for the 'Only etched' case for Str-2, and for other samples lower bound is comparatively higher.

*3.5. Aging properties*

Finally, we have checked the temporal stability of the passivation. We have repeated the measurements after two months of the first measurement and presented in the Fig. 8, which clearly shows that the sulphur passivation and $Al_2O_3$ encapsulation have a noticeable impact on the aging of the samples; for 'Only etched' samples, we observed a large change of the linewidth and the standard deviation after the above said period, whereas the linewidth increment and the degradation of the surface was relatively slower for 'Passivated' samples. We can also notice that the degradation corresponding to a thicker encapsulating oxide of 12 nm is slower than the 6 nm one.

## 4. Conclusions

In conclusion, we have shown that the sulphur passivation of the GaAs surface followed by a dielectric layer of $Al_2O_3$ of 6–12 nm results in the partial recovery of the linewidth of the excitonic emission for dot-to-surface distances of ≤40 nm, which indicates a decrease of surface states density at passivated surface. The illustrated results are of importance because of the possible applications in the different nanostructures (pillar-like photonic cavities, metal-semiconductor-metal antenna, bull's eye cavity, nanocone cavity etc.), where wet or dry etching is involved or the surface to dot distance is quite small without any etching involvement. The charge fluctuation strength, the main reason for spectral diffusion, can be lessened by the proposed passivation scheme. Furthermore, the described passivation process can recover the PL intensity for single GaAs QDs. Noticeably the surface passivation can only decrease the density of surface states and so also the strength of random charge fluctuation but cannot change anything in the bulk region. So, for the quantum dots well below the surface, the charge noise arising from only the bulk volume cannot be suppressed without optimizing the quality of the growth and/or applying external electric fields [50].

**Acknowledgments**

The authors would like to thank Y. Zakharko, T. Krieger, X. Yuan, and S. Panda for the fruitful discussions. This project has received funding from the European Union's Horizon 2020 Research and Innovation Program under the European Union's Horizon 2020 Research and Innovation Programme (SPQRel; 679183). This work has been also supported by the Austrian Science Fund (FWF): P29603, the Linz Institute of Technology (LIT), and the LIT Secure and Correct Systems Lab, financed by the state of Upper Austria.

**Author contributions**

**Santanu Manna:** Conceptualization, Investigation, Methodology, Software, Formal analysis, Data Curation, Writing - Original Draft, Visualization, Validation, Supervision. **Huiying Huang:** Investigation, Methodology, Writing - Review & Editing, Visualization. **Saimon Filipe Covre da Silva:** Investigation, Writing - Review & Editing. **Christian Schimpf:** Methodology, Writing - Review & Editing. **Michele B. Rota:** Investigation, Writing - Review & Editing. **Barbara Lehner:** Software, Methodology. **Marcus Reindl:** Methodology, Writing - Review & Editing. **Rinaldo Trotta:** Conceptualization, Resources, Project administration, Funding acquisition, Writing - Review & Editing. **Armando Rastelli:** Conceptualization, Resources, Project administration, Funding acquisition, Writing - Review & Editing

**References**

[1]   M.J.A. Schütz, Quantum dots for quantum information processing: Controlling and Exploiting the Quantum Dot Environment, Ludwig-Maximilians-Universität München, 2015.
[2]   P. Michler, Single quantum dots fundamentals, Springer-Verlag Berlin Heidelberg New York, 2003.
[3]   A. Faraon, A. Majumdar, D. Englund, E. Kim, M. Bajcsy, J. Vučković, Integrated quantum optical networks based on quantum dots and photonic crystals, New J. Phys. 13 (2011). https://doi.org/10.1088/1367-2630/13/5/055025.
[4]   C. Santori, D. Fattal, Y. Yamamoto, Single-photon Devices and Applications, Wiley, 2010.
[5]   A.J. Shields, Semiconductor quantum light sources, Nat. Photonics. 1 (2007) 215–223. https://doi.org/10.1038/nphoton.2007.46.
[6]   F. Basso Basset, M.B. Rota, C. Schimpf, D. Tedeschi, K.D. Zeuner, S.F. Covre Da Silva, M. Reindl, V. Zwiller, K.D. Jöns, A. Rastelli, R. Trotta, Entanglement Swapping with Photons Generated on Demand by a Quantum Dot, Phys. Rev. Lett. 123 (2019) 160501. https://doi.org/10.1103/PhysRevLett.123.160501.
[7]   M. Reindl, D. Huber, C. Schimpf, S.F. Covre da Silva, M.B. Rota, H. Huang, V. Zwiller, K.D. Jöns, A. Rastelli, R. Trotta, All-photonic quantum teleportation using on-demand solid-state quantum emitters, Sci. Adv. 4 (2018) eaau1255. https://doi.org/10.1126/sciadv.aau1255.
[8]   P. Tighineanu, C.L. Dreeßen, C. Flindt, P. Lodahl, A.S. Sørensen, Phonon Decoherence of Quantum Dots in Photonic Structures: Broadening of the Zero-Phonon Line and the Role of Dimensionality, Phys. Rev. Lett. 120 (2018). https://doi.org/10.1103/PhysRevLett.120.257401.
[9]   A. Majumdar, E.D. Kim, J. Vučković, Effect of photogenerated carriers on the spectral diffusion of a quantum dot coupled to a photonic crystal cavity, Phys. Rev. B. 84 (2011) 1–6. https://doi.org/10.1103/PhysRevB.84.195304.
[10]  A. V. Khaetskii, D. Loss, L. Glazman, Electron spin decoherence in quantum dots due to interaction with nuclei, Phys. Rev. Lett. 88 (2002) 186802. https://doi.org/10.1103/PhysRevLett.88.186802.
[11]  J. Liu, K. Konthasinghe, M. Davanço, J. Lawall, V. Anant, V. Verma, R. Mirin, S.W. Nam, J.D. Song, B. Ma, Z.S. Chen, H.Q. Ni, Z.C. Niu, K. Srinivasan, Single Self-Assembled InAs/GaAs Quantum Dots in Photonic Nanostructures: The Role of Nanofabrication, Phys. Rev. Appl. 9 (2018) 64019. https://doi.org/10.1103/PhysRevApplied.9.064019.
[12]  C.F. Wang, A. Badolato, I. Wilson-Rae, P.M. Petroff, E. Hu, J. Urayama, A. Imamoğlu, Optical properties of single InAs quantum dots in close proximity to surfaces, Appl. Phys. Lett. 85 (2004) 3423–3425. https://doi.org/10.1063/1.1806251.
[13]  S.M. Sze, K.K. Ng, Physics of Semiconductor Devices, John Wiley & Sons, Inc., Hoboken, NJ, USA, 2006. https://doi.org/10.1002/0470068329.
[14]  Y.L. Chang, I.H. Tan, Y.H. Zhang, D. Bimberg, J. Merz, E. Hu, Reduced quantum efficiency of a near-surface quantum well, J. Appl. Phys. 74 (1993) 5144–5148. https://doi.org/10.1063/1.354276.
[15]  S. Checcucci, P. Lombardi, S. Rizvi, F. Sgrignuoli, N. Gruhler, F.B. Dieleman, F. S Cataliotti, W.H. Pernice, M. Agio, C. Toninelli, Beaming light from a quantum emitter with a planar optical antenna, Light Sci. Appl. 6 (2017).

https://doi.org/10.1038/lsa.2016.245.

[16] H. Galal, M. Agio, Highly efficient light extraction and directional emission from large refractive-index materials with a planar Yagi-Uda antenna, Opt. Mater. Express. 7 (2017) 1634. https://doi.org/10.1364/ome.7.001634.

[17] Y.A. Kelaita, K.A. Fischer, T.M. Babinec, K.G. Lagoudakis, T. Sarmiento, A. Rundquist, A. Majumdar, J. Vučković, Hybrid metal-dielectric nanocavity for enhanced light-matter interactions, Opt. Mater. Express. 7 (2017) 231. https://doi.org/10.1364/ome.7.000231.

[18] X. Liu, T. Asano, S. Odashima, H. Nakajima, H. Kumano, I. Suemune, Bright single-photon source based on an InAs quantum dot in a silver-embedded nanocone structure, Appl. Phys. Lett. 102 (2013) 131114. https://doi.org/10.1063/1.4801334.

[19] J. Liu, R. Su, Y. Wei, B. Yao, S.F.C. da Silva, Y. Yu, J. Iles-Smith, K. Srinivasan, A. Rastelli, J. Li, X. Wang, A solid-state source of strongly entangled photon pairs with high brightness and indistinguishability, Nat. Nanotechnol. 14 (2019) 586–593. https://doi.org/10.1038/s41565-019-0435-9.

[20] H. Wang, H. Hu, T.H. Chung, J. Qin, X. Yang, J.P. Li, R.Z. Liu, H.S. Zhong, Y.M. He, X. Ding, Y.H. Deng, Q. Dai, Y.H. Huo, S. Höfling, C.Y. Lu, J.W. Pan, On-Demand Semiconductor Source of Entangled Photons Which Simultaneously Has High Fidelity, Efficiency, and Indistinguishability, Phys. Rev. Lett. 122 (2019). https://doi.org/10.1103/PhysRevLett.122.113602.

[21] L. Sapienza, M. Davanço, A. Badolato, K. Srinivasan, Nanoscale optical positioning of single quantum dots for bright and pure single-photon emission, Nat. Commun. 6 (2015). https://doi.org/10.1038/ncomms8833.

[22] L. Zhou, B. Bo, X. Yan, C. Wang, Y. Chi, X. Yang, Brief review of surface passivation on III-V semiconductor, Crystals. 8 (2018) 1–14. https://doi.org/10.3390/cryst8050226.

[23] T. Gougousi, Atomic layer deposition of high-k dielectrics on III–V semiconductor surfaces, Prog. Cryst. Growth Charact. Mater. 62 (2016) 1–21. https://doi.org/10.1016/j.pcrysgrow.2016.11.001.

[24] H. Hasegawa, M. Akazawa, Surface passivation technology for III-V semiconductor nanoelectronics, Appl. Surf. Sci. (2008). https://doi.org/10.1016/j.apsusc.2008.07.002.

[25] M. Passlack, M. Hong, R.L. Opila, J.P. Mannaerts, J.R. Kwo, GaAs surface passivation using in-situ oxide deposition, Appl. Surf. Sci. 104–105 (1996) 441–447. https://doi.org/10.1016/S0169-4332(96)00184-5.

[26] R. Krishnamurthy, Passivation of GaAs and GaInAsP Semiconducting Materials, (1998).

[27] H. Oigawa, J.F. Fan, Y. Nannichi, H. Sugahara, M. Oshima, Universal passivation effect of (Nh4)2sx treatment on the surface of iii-v compound semiconductors, Jpn. J. Appl. Phys. 30 (1991) L322–L325. https://doi.org/10.1143/JJAP.30.L322.

[28] J. Shin, K.M. Geib, C.W. Wilmsen, Z. Lilliental-Weber, The chemistry of sulfur passivation of GaAs surfaces, J. Vac. Sci. Technol. A Vacuum, Surfaces, Film. 8 (1990) 1894–1898. https://doi.org/10.1116/1.576822.

[29] B. Guha, F. Marsault, F. Cadiz, L. Morgenroth, V. Ulin, V. Berkovitz, A. Lemaître, C. Gomez, A. Amo, S. Combrié, B. Gérard, G. Leo, I. Favero, Surface-enhanced gallium arsenide photonic resonator with quality factor of $6 \times 10^6$, Optica. 4 (2017) 218. https://doi.org/10.1364/optica.4.000218.

[30] C. Headley, L. Fu, P. Parkinson, X. Xu, J. Lloyd-Hughes, C. Jagadish, M.B. Johnston, Improved performance of GaAs-based terahertz emitters via surface passivation and silicon nitride encapsulation, IEEE J. Sel. Top. Quantum Electron. 17 (2011) 17–21.

https://doi.org/10.1109/JSTQE.2010.2047006.

[31] K. Adlkofer, E.F. Duijs, F. Findeis, M. Bichler, A. Zrenner, E. Sackmann, G. Abstreiter, M. Tanaka, Enhancement of photoluminescence from near-surface quantum dots by suppression of surface state density, Phys. Chem. Chem. Phys. 4 (2002) 785–790. https://doi.org/10.1039/b108683a.

[32] J. Martín-Sánchez, Y. González, P. Alonso-González, L. González, Improvement of InAs quantum dots optical properties in close proximity to GaAs(0 0 1) substrate surface, J. Cryst. Growth. 310 (2008) 4676–4680. https://doi.org/10.1016/j.jcrysgro.2008.08.041.

[33] D. Huber, M. Reindl, J. Aberl, A. Rastelli, R. Trotta, Semiconductor quantum dots as an ideal source of polarization-entangled photon pairs on-demand: A review, J. Opt. 20 (2018). https://doi.org/10.1088/2040-8986/aac4c4.

[34] C. Heyn, D. Sonnenberg, W. Hansen, Local Droplet Etching: Self-assembled Nanoholes for Quantum Dots and Nanopillars, in: 2013: pp. 363–384. https://doi.org/10.1007/978-1-4614-9472-0_15.

[35] Y.H. Huo, A. Rastelli, O.G. Schmidt, Ultra-small excitonic fine structure splitting in highly symmetric quantum dots on GaAs (001) substrate, Appl. Phys. Lett. 102 (2013) 152105. https://doi.org/10.1063/1.4802088.

[36] D. Huber, M. Reindl, S.F. Covre Da Silva, C. Schimpf, J. Martín-Sánchez, H. Huang, G. Piredda, J. Edlinger, A. Rastelli, R. Trotta, Strain-Tunable GaAs Quantum Dot: A Nearly Dephasing-Free Source of Entangled Photon Pairs on Demand, Phys. Rev. Lett. 121 (2018) 33902. https://doi.org/10.1103/PhysRevLett.121.033902.

[37] J.H. Kim, D.H. Lim, G.M. Yang, Selective etching of AlGaAs/GaAs structures using the solutions of citric acid/H2O2 and de-ionized H2O/buffered oxide etch, J. Vac. Sci. Technol. B Microelectron. Nanom. Struct. 16 (1998) 558–560. https://doi.org/10.1116/1.589862.

[38] P. Kumar, S. Kanakaraju, D.L. Devoe, Sacrificial etching of AlxGa1-xAs for III-V MEMS surface micromachining, Appl. Phys. A Mater. Sci. Process. 88 (2007) 711–714. https://doi.org/10.1007/s00339-007-4032-7.

[39] X. Yi, L. Hung-Chun, P. Ye, Simplified Surface Preparation for GaAs Passivation Using Atomic Layer-Deposited High-k Dielectrics, IEEE Trans. Electron Devices. 54 (2007) 1811–1817.

[40] C.F. Yen, M.K. Lee, J.C. Lee, Electrical characteristics of Al2O3/TiO2/Al2O3 prepared by atomic layer deposition on (NH4)2S-treated GaAs, Solid. State. Electron. 92 (2014) 1–4. https://doi.org/10.1016/j.sse.2013.10.002.

[41] X.Y. Hou, W.Z. Cai, Z.Q. He, P.H. Hao, Z.S. Li, X.M. Ding, X. Wang, Electrochemical sulfur passivation of GaAs, Appl. Phys. Lett. 60 (1992) 2252–2254. https://doi.org/10.1063/1.107475.

[42] W. Mönch, Electronic Structure of Metal-Semiconductor Contacts, Springer Netherlands, 1990.

[43] J. Schreiber, S. Hildebrandt, W. Kircher, T. Richter, Optical investigations of surface and interface properties at III-V semiconductors, Mater. Sci. Eng. B. 9 (1991) 31–35. https://doi.org/10.1016/0921-5107(91)90144-K.

[44] A. Berthelot, I. Favero, G. Cassabois, C. Voisin, C. Delalande, P. Roussignol, R. Ferreira, J.M. Gérard, Unconventional motional narrowing in the optical spectrum of a semiconductor quantum dot, Nat. Phys. 2 (2006) 759–764. https://doi.org/10.1038/nphys433.

[45] A. V. Kuhlmann, J. Houel, A. Ludwig, L. Greuter, D. Reuter, A.D. Wieck, M. Poggio, R.J. Warburton, Charge noise and spin noise in a semiconductor quantum device, Nat. Phys. 9 (2013) 570–575. https://doi.org/10.1038/nphys2688.

[46] C. Schimpf, M. Reindl, P. Klenovský, T. Fromherz, S.F. Covre Da Silva, J. Hofer, C. Schneider, S. Höfling, R. Trotta, A. Rastelli, Resolving the temporal evolution of line broadening in single quantum emitters, Opt. Express. 27 (2019) 35290. https://doi.org/10.1364/oe.27.035290.

[47] R. Kubo, A Stochastic Theory of Line Shape, Stoch. Process. Chem. Phys. 15 (2006) 101–127. https://doi.org/10.1002/9780470143605.ch6.

[48] P.W. Anderson, A Mathematical Model for the Narrowing of Spectral Lines by Exchange or Motion, J. Phys. Soc. Japan. 9 (1954) 316–339. https://doi.org/10.1143/JPSJ.9.316.

[49] D. V. Matyushov, Kubo's Line Shape Function for a Linear-Quadratic Chromophore-Solvent Coupling, J. Phys. Chem. B. 119 (2015) 9006–9008. https://doi.org/10.1021/jp5081059.

[50] L. Zhai, M.C. Löbl, G.N. Nguyen, J. Ritzmann, A. Javadi, C. Spinnler, A.D. Wieck, A. Ludwig, R.J. Warburton, Low-Noise GaAs Quantum Dots for Quantum Photonics, (2020).

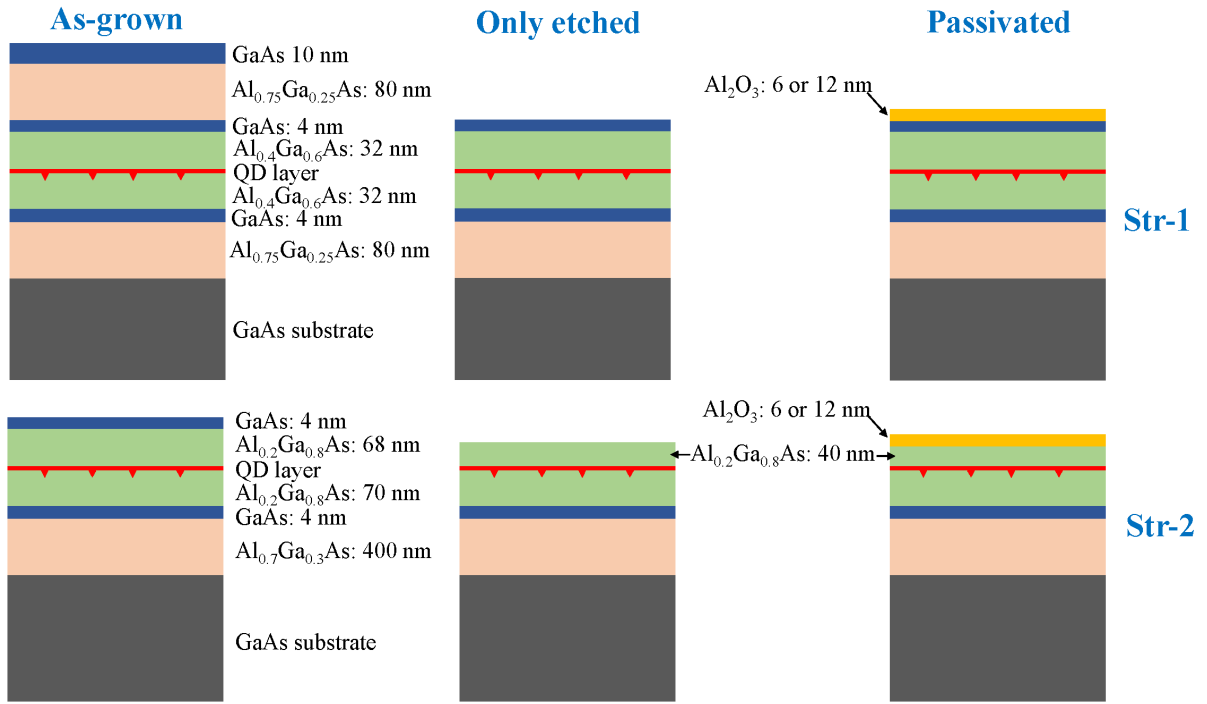

**Fig. 1.** Schematic diagram of the two different sample structures (namely, Str-1 and Str-2) corresponding to 'As-grown', 'Only etched' and 'Passivated' samples with 6 or 12 nm $Al_2O_3$ encapsulation. No GaAs cap layer is present for Str-2 unlike Str-1 in case of etching and passivation.

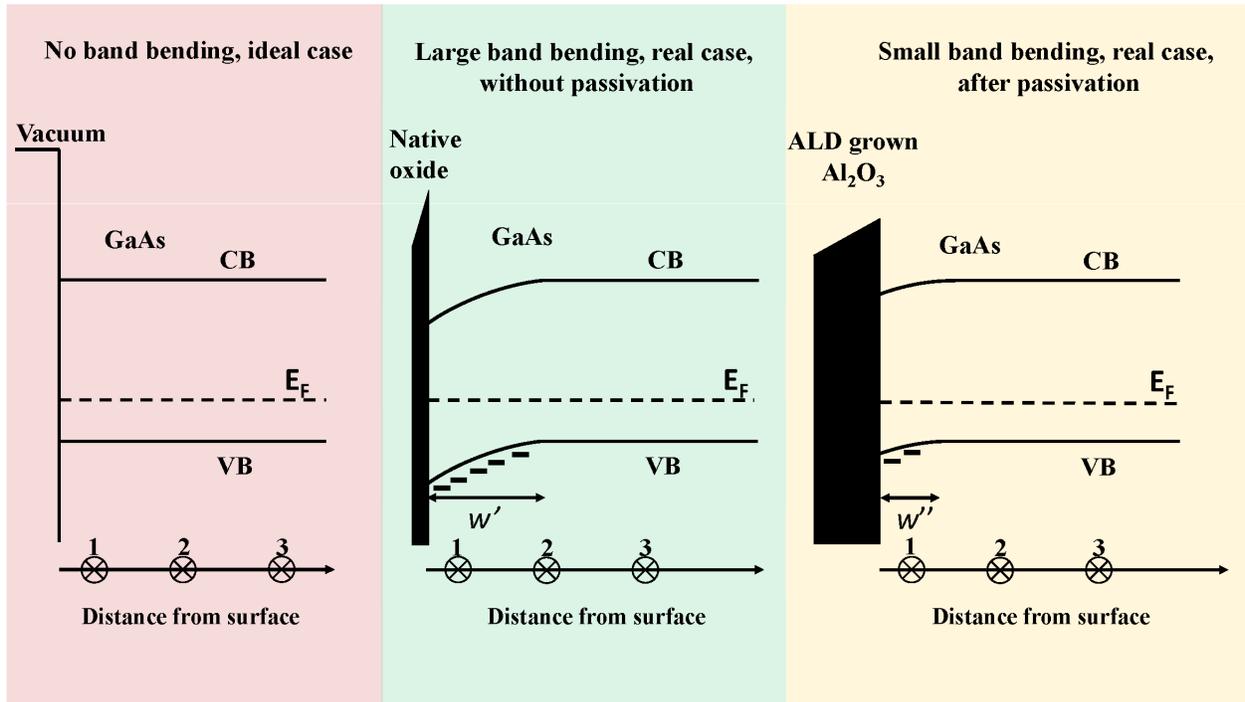

**Fig. 2.** Simplified schematic band diagrams indicate the band bending in presence and absence of surface states. The built-in electric field stemming from surface states affect the QDs depending on the extent of the depletion region and QD depth from the surface. w′ and w″ denote depletion widths for large and small band bending cases, respectively, with the condition w′> w″.

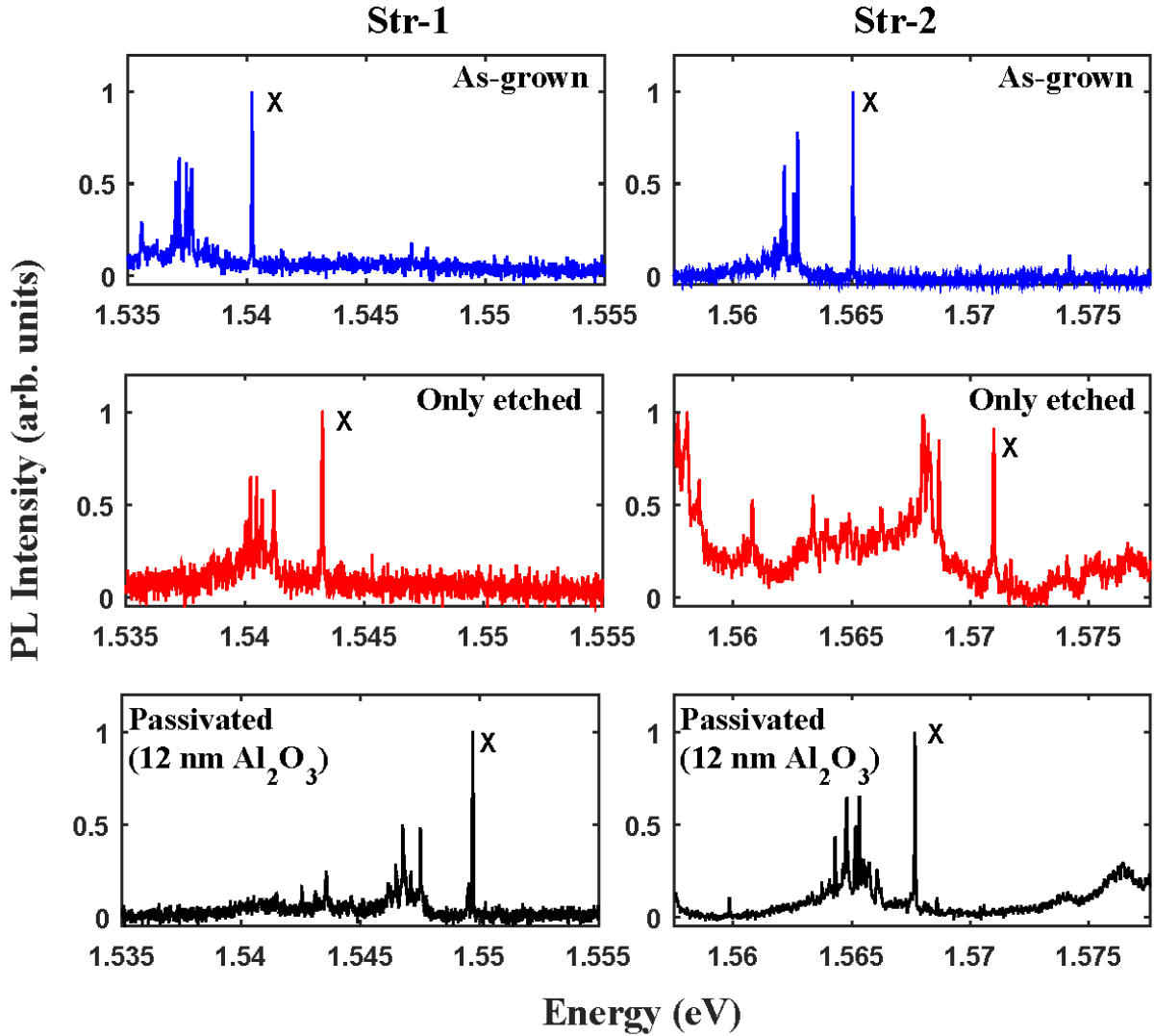

**Fig. 3.** Typical single quantum dot spectra from GaAs QDs grown by droplet etching epitaxy are presented. Different spectra were collected for the different samples, see Fig. 1. PL background modulation was severe for 'Only etched' case for Str-2 compared to other cases due to the absence of GaAs cap layer.

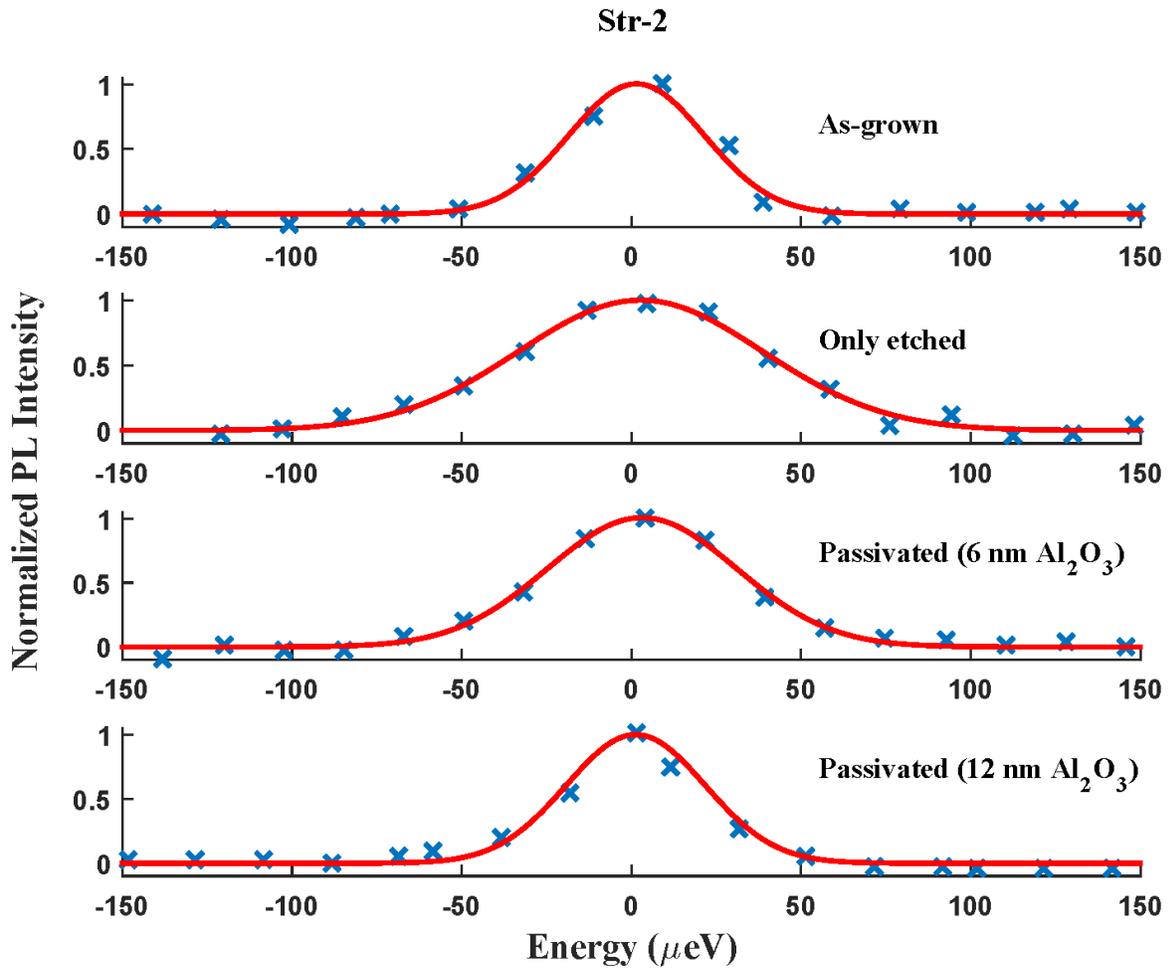

**Fig. 4.** Photoluminescence spectra of representative neutral excitons confined in GaAs QDs in Str-2 samples (see Fig. 1) and their corresponding Gaussian fits. The spectra were shifted along the energy axis to facilitate the comparison.

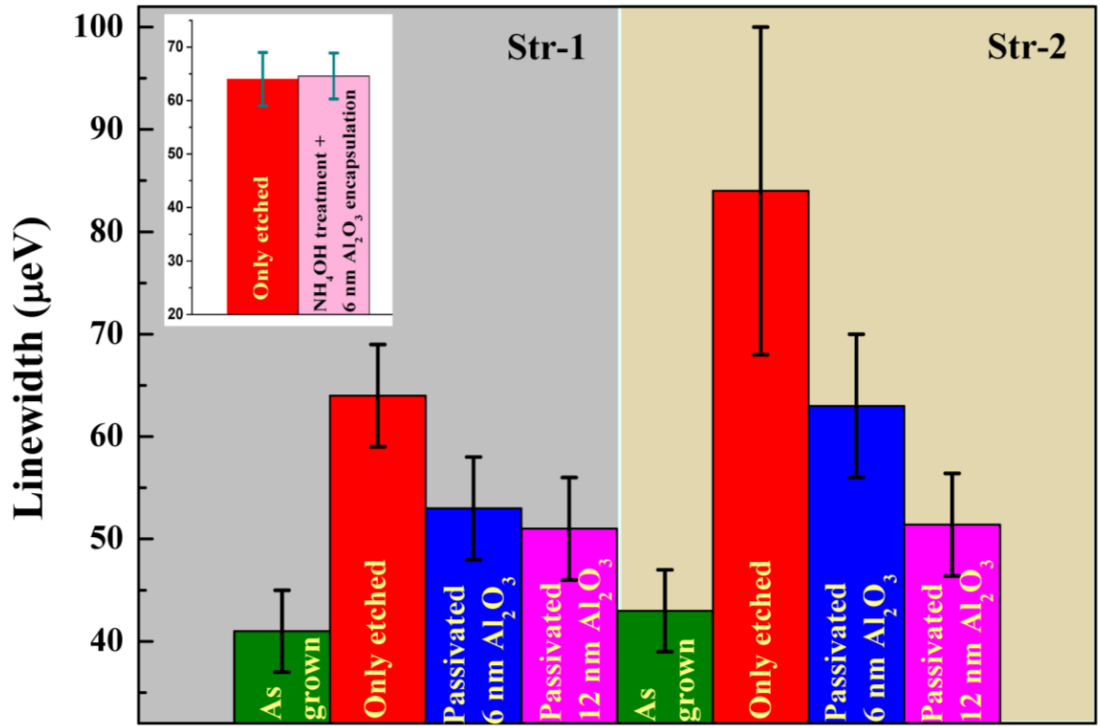

**Fig. 5.** Bar plots corresponding to the statistical measurements of linewidths of neutral excitons confined in GaAs QDs in both structures Str-1 and Str-2 (see Fig. 1). Inset shows the comparison between 'Only etched' case and one with 'NH$_4$OH treatment + 6 nm Al$_2$O$_3$ encapsulation' in case of Str-1.

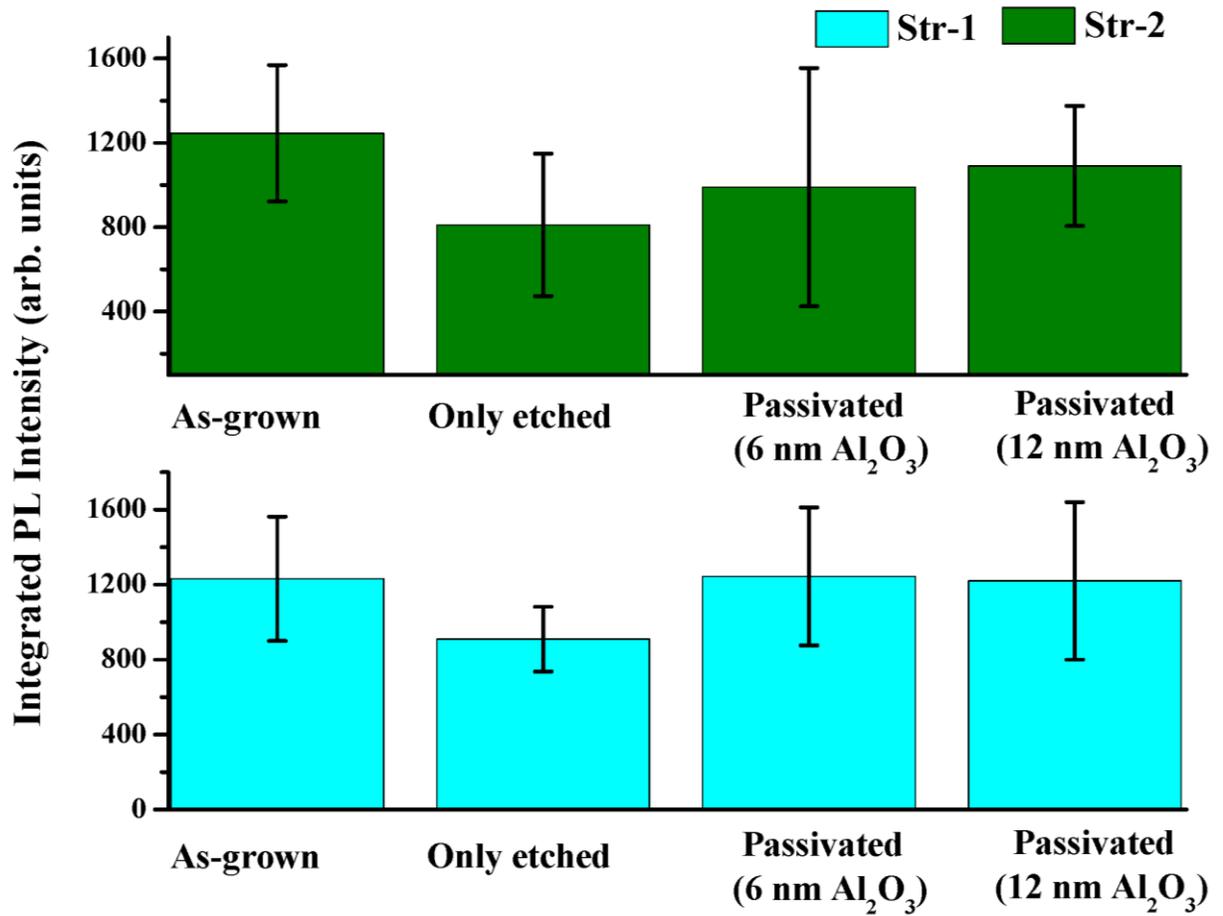

**Fig. 6.** Bar plots corresponding to the statistical measurements of PL integrated intensity of neutral exciton line for different samples (both Str-1 and Str-2).

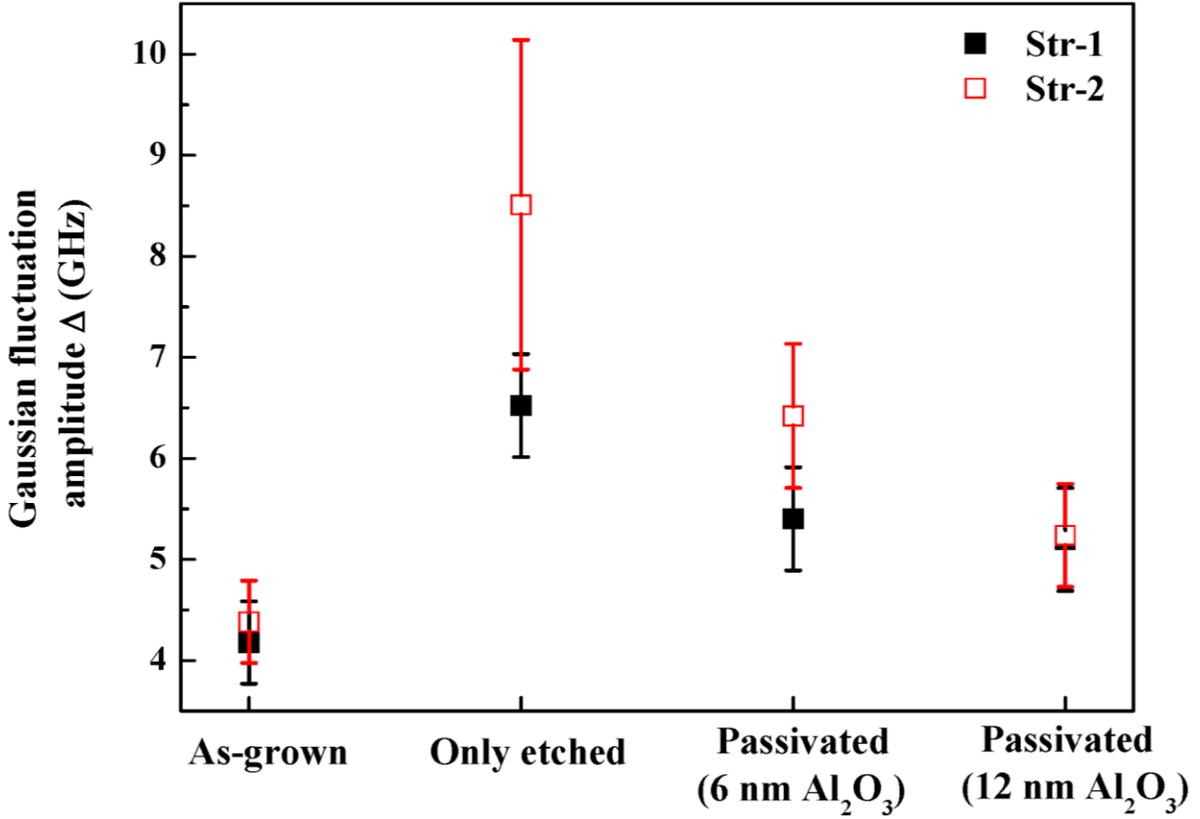

**Fig. 7.** Strengths or amplitudes of the random Gaussian fluctuation (trapping/detrapping of charge carriers) are estimated from neutral exciton linewidth using the Kubo-Anderson model for different samples (both Str-1 and Str-2).

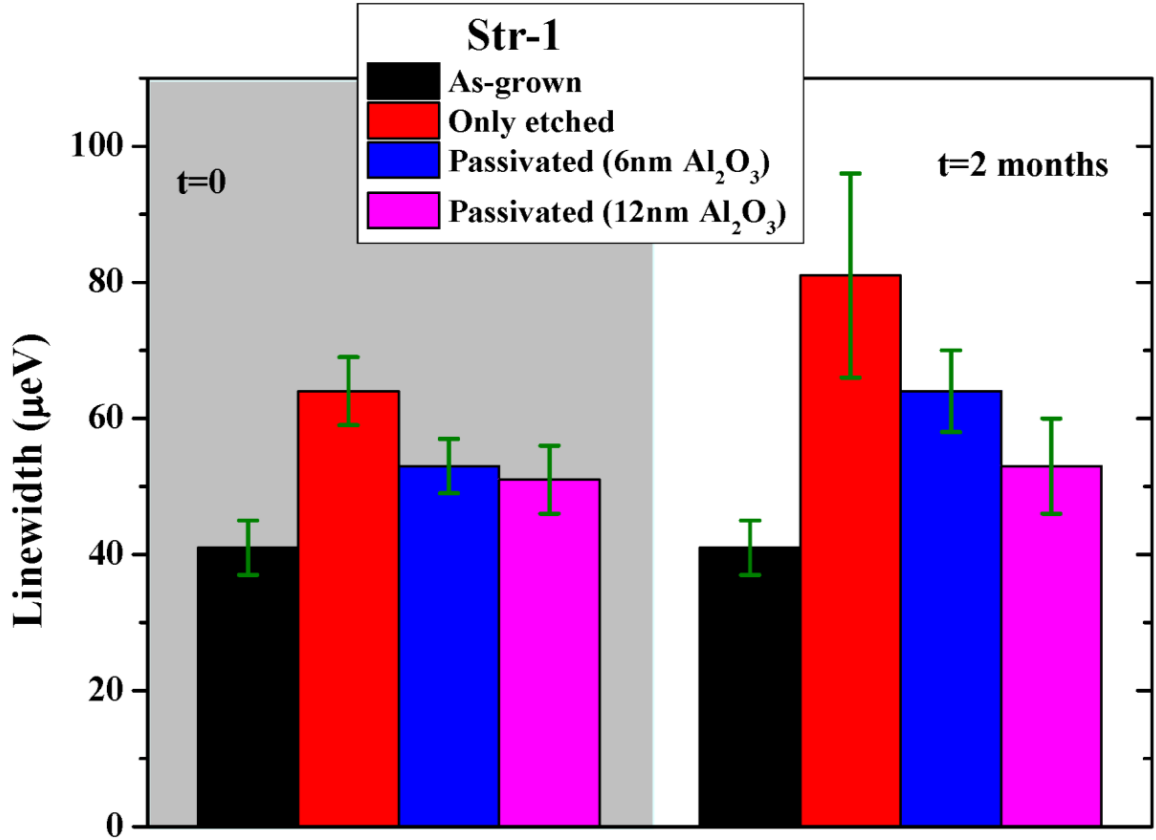

**Fig. 8.** Aging of the samples for Str-1 indicates that degradation of the linewidth happens over time for all cases. For 'Only etched' case, the degradation is faster compared to the one with sulphur 'Passivated' cases. Also, samples with a thicker $Al_2O_3$ encapsulation degrade slower.